\documentclass[prd,twocolumn,superscriptaddress,floatfix,showpacs]{revtex4}%
\usepackage{amsfonts}
\usepackage{amsmath}
\usepackage{amssymb}
\usepackage{xcolor}
\usepackage{graphicx}
\font\mybb=msbm10 at 10pt
\def\bb#1{\hbox{\mybb#1}}
\def\be{\begin{equation}}
\def\ee{\end{equation}}
\def\bea{\begin{eqnarray}}
\def\eea{\end{eqnarray}}

\begin{document}
\title{Complete nonlinear action for supersymmetric multiple D$0$-brane system}

\author{Igor Bandos}
%\email{igor.bandos@ehu.eus}
\affiliation{Department of Physics and EHU Quantum Center, University of the Basque Country UPV/EHU,
P.O. Box 644, 48080 Bilbao, Spain,}
\affiliation{IKERBASQUE, Basque Foundation for Science,
48011, Bilbao, Spain, }
%\affiliation{
%EHU Quantum Center, Universidad del País Vasco/Euskal Herriko Unibertsitatea, UPV/EHU,
%Barrio Sarriena, s/n, 48940 Leioa, Spain
%}

\author{Unai D.M. Sarraga$^1$}
%\affiliation{ University of the Basque Country UPV/EHU,
%P.O. Box 644, 48080 Bilbao, Spain}

\begin{abstract}

We present a complete nonlinear action for the dynamical system of nearly coincident multiple D$0$-branes (mD$0$) which possesses, besides manifest spacetime (target superspace) supersymmetry, also the  worldline supersymmetry, a counterpart of the local fermionic $\kappa$-symmetry of single D$0$-brane (Dirichlet superparticle). The action contains an arbitrary non-vanishing function ${\cal M}({\cal H})$ of the relative motion Hamiltonian ${\cal H}$. The ten-dimensional ($D=10$) mD$0$ model with particular form of ${\cal M}({\cal H})$ can be obtained by dimensional reduction from the action of $D=11$ multiple M-wave (mM$0$) system.

\end{abstract}


\maketitle

\setcounter{equation}{0}

\section{Introduction}
Dirichlet $p$-branes or D$p$-branes \footnote{Of these D1-branes are  Dirichlet strings or D-stings, D2-branes are  Dirichlet membranes and D$0$-branes are so-called Dirichlet particles, massive supersymmetric particles. The set of higher $p$-branes contains the maximal D$9$-brane which is spacetime filling as the String theory is 10 dimensional. In the original language of string model \cite{Green:1987sp} the string ending on D$9$-brane is called the superstring with free ends.} are the  supersymmetric extended objects on which the fundamental $D=10$ superstring can have its ends attached \cite{Dai:1989ua,Horava:1989ga}. Their especially important role in String Theory \cite{Green:1987sp} was appreciated after the famous paper by J. Polchinski \cite{Polchinski:1995mt} where it was shown that they carry nontrivial charges with respect to Ramond-Ramond (RR) fields (see \cite{Johnson:2003glb} for a comprehensive review).

The worldvolume action for single  super-D$p$-brane is known
\cite{Howe:1996mx,Cederwall:1996pv,Aganagic:1996pe,Cederwall:1996ri,Aganagic:1996nn,Bergshoeff:1996tu,Bandos:1997rq}
to be given by the sum of supersymmetrised Dirac-Born-Infeld (DBI) term and a Wess-Zumino term describing the coupling to RR fields. Both terms contain the field strength of $d=(p+1)$ dimensional worldvolume gauge field and in the weak field limit, after fixing the static gauge the first DBI term reduces to the action of the supersymmetric  Abelian gauge field theory. Also the Wess--Zumino term in this gauge is expressed through the fields of Abelian super-Yang-Mills multiplet.

The quest for an effective action for the multiple D$p$-brane system, i.e. the system of $N$ nearly coincident D$p$-branes and strings ending on these D$p$-branes,  can be followed back to the seminal paper by E. Witten \cite{Witten:1995im} where he argued that the gauge fixed description of its weak field limit is given by the non-Abelian U($N$) super-Yang-Mills (SYM) action. Despite a number of very interesting results
obtained during the passed 26 years \cite{Sorokin:2001av,Drummond:2002kg,Janssen:2002vb,Janssen:2002cf,Panda:2003dj,Janssen:2003ri,Lozano:2005kf,Howe:2005jz,Howe:2006rv,Howe:2007eb,Bandos:2018ntt,Bandos:2021vrq}
the complete nonlinear supersymmetric action for the dynamical system of multiple D$p$-branes (mD$p$) is not known presently even for the simplest case of $p=0$ \footnote{Two comments should be made in this respect. First this statement refers to the mD$p$ action in its form similar to the action of single D$p$-brane and which has the weak field limit described in \cite{Witten:1995im}, so it does not apply to a very interesting construction on `-1 quantization level' proposed and elaborated in \cite{Howe:2005jz,Howe:2006rv,Howe:2007eb} (see \cite{Bandos:2018ntt} for more discussion). Also notice  an action proposed in \cite{Bandos:2018ntt}  which will appear as a particular (simplest) case of nonlinear actions presented in this work.}.

In this paper we present a nonlinear action which possesses several properties expected from the action of mD$0$ system. Particularly,  it is manifestly invariant under Poincar\'e symmetry, SU(N) gauge symmetry and spacetime (type IIA target superspace) supersymmetry, and also possesses local worldline supersymmetry generalizing the $\kappa$-symmetry of single D$0$-brane (massive type II $D=10 $ superparticle) action \footnote{The $\kappa$-symmetry was discovered for massive superparticle in
\cite{deAzcarraga:1982dhu,deAzcarraga:1982njd} and for massless one in \cite{Siegel:1983hh}. The identification of  $\kappa$-symmetry with worldline supersymmetry was established in \cite{Sorokin:1988jor} , see \cite{Sorokin:1989jj} for review.
}. This latter fact is especially important because it guarantees that the ground state of this dynamical system is supersymmetric which is expected in the case of multiple D$0$-brane system.

The rest of the paper is organized as follows. In sec. II we present the complete supersymmetric and nonlinear candidate action for multiple D$0$-brane system. The rigid spacetime supersymmetry and local worldsheet supersymmetry transformations leaving this action invariant are described in sec. III. The technical details on the derivation of these results can be found in Appendix D which uses the approach and ingredients described in Appendices A-C. Sec. IV contains our conclusions and  discussion of the results.

\section{Supersymmetric  nonlinear action }

The nonlinear action which we have found is written in terms of center of energy variables of mD$0$ system,
which are the same as in the case of single D$0$-brane, and matrix variables describing the relative motion of mD$0$ constituents.
The set of center of energy variables contains  coordinate functions describing the embedding of the center of energy worldline in flat  type IIA superspace, bosonic 10-vector and two fermionic Majorana-Weyl spinors
\be
Z^M(\tau)= (x^\mu(\tau),\theta^{1 \alpha}(\tau), \theta^2_{\alpha}(\tau) )~, \quad
\ee
$\mu=0,...,9$, $\alpha=1,...,16$, as well as the spinor moving frame variables which we will describe below.
 The relative motion variables are matrix fields  from the 1d extended (${\cal N}=16$) SU($N$) SYM multiplet, the set of which can be split on matter fields,  9$+$9 bosonic and 16 fermionic Hermitean traceless $N\times N$ matrix fields
\be\label{matter}
{\bb X}^i(\tau) , \quad {\bb P}^i(\tau) , \quad {\mathbf{\Psi}}_q (\tau) , \qquad \ee
$i=1,...,9, \quad q=1,...,16$,
and the bosonic anti-Hermitean traceless $N\times N$ matrix 1-form
\be\label{bfA=}
{\bb A}= \text{d}\tau {\bb A}_\tau (\tau)
\ee
containing the $su(N)$ valued worldline  gauge field  ${\bb A}_\tau (\tau)$.
Besides SU($N$) gauge transformations, the matrix fields are transformed by local SO$(9)$ transformations according to their vector and spinor indices $i=1,...,9$ and $q=1,...,16$. These will also act on spinor frame variables and describe the gauge symmetry of the mD$0$ action.

The action has the form
\begin{eqnarray}
\label{SmD0=} && S_{\text{mD0}} = m \int_{\mathcal{W}^1} {E}^{0}     -im  \int_{\mathcal{W}^1} (\text{d}\theta^1\theta^2-\theta^1 \text{d}\theta^2)
    + \quad \nonumber \\ && +  \frac 1 {\mu^6} \int_{\mathcal{W}^1}   \left(\text{tr}\left({\bb P}^i \text{D} {\bb X}^i + 4i {\mathbf{ \Psi}}_q \text{D}
{\mathbf{ \Psi}}_q  \right) +  \frac 2 {\cal M} {E}^{0}\, {\cal H}\right) \nonumber \\ && -  \frac 1 {\mu^6} \int_{\mathcal{W}^1} \frac {\text{d} {\cal M}}{ {\cal M} }
 {\rm tr} ({\bb P}^i{\bb X}^i) +   \frac 1 {\mu^6}  \int_{\mathcal{W}^1}
  \frac 1 {\sqrt{2{\cal M}}}({E}{}^{1q}-{E}{}^{2}_{q}) \times \nonumber \\ && {} \qquad \times {\rm tr}\left(-4i (\gamma^i {\mathbf{ \Psi}})_q  {\bb P}^i  + {1\over 2}
(\gamma^{ij} {\mathbf{ \Psi}})_q  [{\bb X}^i, {\bb X}^j]  \right)  \qquad
  \end{eqnarray}
where  $m$ and $\mu$ are constants of dimension of mass and
\begin{eqnarray}
\label{HmM0==} {\cal H} &=& {1\over 2} \text{tr}\left( {\bb P}^i {\bb P}^i \right)   - {1\over 64}
\text{tr}\left[ {\bb X}^i ,{\bb X}^j \right]^2  - 2\,  \text{tr}\left({\bb X}^i\, {\bf \Psi}\gamma^i {\bf \Psi}\right)  \quad
  \end{eqnarray}
has the meaning of the relative motion Hamiltonian.

Actually the first line of \eqref{SmD0=}  formally coincides with the action of single D$0$-brane, i.e. massive $D=10$ type IIA superparticle in its moving frame formulation \cite{Bandos:2018ntt,Bandos:2000tg} (see below for the description of $E^{0}$ in it and Appendix B for some details). In this case $m$ plays the role of the superparticle mass. In contrast, the constant $\mu$ characterizes the interaction of the center of energy and relative motion sector as well as the self-interaction of this latter. Notice that to simplify and to make more transparent the dependence of the action on this parameter we have chosen non-canonical dimensions for the matrix matter fields \eqref{matter}.
In particular, with this choice of dimensions of matrix fields, the relative motion Hamiltonian ${\cal H}$ \eqref{HmM0==} is $\mu$-independent. However its dimension becomes (mass$^6$) so that
${\cal H}/\mu^6$ is dimensionless.

${\cal M}$ in \eqref{SmD0=} is {\it an arbitrary} nonvanishing {\it function} of this dimensionless combination of the relative motion Hamiltonian and coupling constant,
\be
{\cal M}= {\cal M}({\cal H}/\mu^6)\; .
\ee
A particular case of the action \eqref{SmD0=} with
\be\label{cM=m+}
{\cal M} = \frac m 2 +  \sqrt{\frac {m^2} {4}+\frac {\cal H} {\mu^6}}\;
\ee
can be obtained by dimensional reduction of the 11D multiple M-wave (multiple M$0$-branes or mM$0$) system action from \cite{Bandos:2012jz,Bandos:2013uoa} similar to dimensional reduction of its $D=4$ counterpart described in
\cite{Bandos:2021vrq}. Another representative of the family \eqref{SmD0=} with ${\cal M} =m$ was studied in \cite{Bandos:2018ntt} where it was noticed that it cannot be obtained by dimensional reduction from 11D mM$0$ action.

Coming back to the first line of \eqref{SmD0=}, in it $E^{0}$  is the projection of (the pull-back of) 10D Volkov-Akulov 1-form
\be\label{E0:=}
E^{0}=\Pi^\mu u_\mu^{0} \; , \qquad
\Pi^{{\mu}}= \text{d}x^{{\mu}} - i\text{d}\theta^1 \sigma^{{\mu}}\theta^1 - i\text{d}\theta^2 \tilde{\sigma}^{{\mu}}\theta^2
\ee
to one of the vector fields, $ u_\mu^{0}(\tau)$, of moving frame attached to the worldline. That is described by Lorentz group valued 10$\times$10 matrix
\be\label{harmU=10}
(u_\mu^{0}, u_\mu^{i})\in \text{SO}(1,9) \qquad
\ee
composed of the moving frame vectors which obey
\be\label{u0u0=1}
u^{\mu 0}u_\mu^{0}=1\; , \qquad u^{\mu 0} u_\mu^{i}=0\; , \qquad u^{\mu i} u_\mu^{j}=-\delta ^{ij}\; .
\ee
The spinor moving frame described by Spin$(1,9)$ valued matrix
\begin{eqnarray}\label{harmV=10}
v_\alpha{}^q \in \text{Spin}(1,9)\;  \qquad
\end{eqnarray}
provides a kind of square root of the above described moving frame in the sense of Cartan-Penrose-like relations
(see Appendix A for more details)
\begin{eqnarray}\label{u0s=vv}
u_\mu^{0} \sigma^\mu_{\alpha\beta}=v_\alpha{}^q v_\beta{}^q \; , \qquad
u_\mu^{i} \sigma^\mu_{\alpha\beta}=v_\alpha{}^q \gamma^i_{qp}v_\beta{}^p \; ,  \qquad \\
v_{\alpha}^q \tilde{\sigma}{}_{\mu}^{\alpha\beta}v_{\beta}^p= u_\mu^{0} \delta_{qp}+u_\mu^{i} \gamma^i_{qp}\; . \qquad
\end{eqnarray}
In distinction to their $D=4$ counterparts (described in \cite{Bandos:1990ji} and e.g. \cite{Bandos:2021vrq}) Eqs. \eqref{u0s=vv} impose strong constraints on the spinor moving frame field $v_\alpha{}^q=v_\alpha{}^q(\tau)$ reducing the number of its components from the original 16$\times$16=256 to $45={\rm dim}(\text{SO}(1,9))$.

This spinor frame matrix field $v_\alpha{}^q(\tau)$ and its inverse $v_q^\alpha(\tau)$ are used to construct the fermionic forms $E^{1q}$ and $E_2^q$
which enter the last term of the action \eqref{SmD0=},
\be\label{Eq1=}  E^{1q}= \text{d}\theta^{1 \alpha}v_\alpha{}^q\; , \qquad  E_{q}^2= \text{d}\theta_{\alpha}^2v_q{}^\alpha\; .  \qquad \ee
The covariant derivatives in the second line of \eqref{SmD0=}
\begin{eqnarray}
\label{DXi=} \text{D}{\bb X}^i  &:=& \text{d}\tau \text{D}_\tau {\bb X}^i :=  \text{d}{\bb X}^i   - \Omega^{ij} {\bb
X}^j+ [{\bb A},    {\bb X}^i] \; , \qquad \\ \label{DPsi:=} \text{D}\Psi_q  &:=&  \text{d}\tau \text{D}_\tau {\mathbf{\Psi}}_q := \text{d}{\mathbf{\Psi}}_q
   -{1\over 4} \Omega^{ij} \gamma^{ij}_{qp} {\mathbf{\Psi}}_p+ [{\bb A},
 {\mathbf{\Psi}}_q ] \; . \qquad
  \end{eqnarray}
contain, beside the SU$(N)$ gauge field \eqref{bfA=}, also the composite SO$(9)$ connection (Cartan form)
  \be\label{Omij=10} {\Omega}^{ij} = u^{{\mu}i}\text{d} u_{{\mu}}^{j}\; . \qquad
\ee

\section{Local worldline supersymmetry}

The action \eqref{SmD0=} is manifestly invariant under the rigid super-Poincar\'{e} supergroup transformations, including spacetime (target 10D IIA superspace) supersymmetry with constant fermionic parameters $\epsilon^{\alpha 1}$ and $\epsilon_\alpha{}^2$ acting nontrivially only on the center of energy variables,
\bea\label{susy=IIA}
 \delta_\epsilon \theta^{1 \alpha}= \epsilon^{\alpha 1} \; , \qquad \delta_\epsilon \theta_{\alpha}^{2}=\epsilon_\alpha{}^2 \; ,  \qquad \delta_\epsilon v_\alpha^q=0~, \nonumber \\ \delta_\epsilon x^\mu =i\theta^1 \sigma^\mu \epsilon^1+i\theta^2 \tilde{\sigma}{}^\mu \epsilon^2\; .
\eea
It is also invariant  under the SU($N$) gauge symmetry acting on the matrix matter fields by its adjoint representation, provided the $su(N)$ valued 1-form
${\bb A}$ transforms as SU$(N)$ connection, as well as under the SO$(9)$ symmetry acting by vector representation on index $i$  of $u_\mu^i$, ${\bb X}^i$, ${\bb P}^i$ and by its spinor representation on index $q$ of ${\mathbf{\Psi}}_q$ and $v_\alpha^q$.

Furthermore the action is invariant under local fermionic worldline supersymmetry parametrized by fermionic function $\kappa^q(\tau)$ carrying  spinor index of SO$(9)$. It acts on the center of energy variables exactly in the same manner as irreducible $\kappa$-symmetry of single D$0$-brane in its spinor moving frame formulation \cite{Bandos:2018ntt,Bandos:2000tg} (hence notation $\kappa^q(\tau)$),
\bea\label{kappa=}
&& \delta_\kappa \theta^{1 \alpha}=  {\kappa^q} v_q^\alpha/{\sqrt{2}} \; , \qquad \delta_\kappa \theta_{\alpha}^{2}= -  {\kappa^q} v_\alpha{}^q/{\sqrt{2}} \; , \qquad \nonumber   \\ && \delta_\kappa x^\mu =i\delta_\kappa\theta^1 \sigma^\mu \theta^1+i\delta_\kappa\theta^2 \tilde{\sigma}{}^\mu \theta^2  \; , \qquad  \nonumber   \\
 && \delta_\kappa v_\alpha^q=0\qquad \Rightarrow \qquad  \delta_\kappa u_{{\mu}}^{0} =0= \delta_\kappa u_{{\mu}}^{i}\; .
 \eea
The action of worldline SUSY on the matrix  fields includes essentially nonlinear terms some of which are proportional to the derivative of the function
${\cal M}$ with respect to its argument and, hence to additional power of $\frac {1}{\mu^6}$,
\be
\delta {\cal M} ({\cal H}/{\mu^6}) = \dfrac {1}{{\mu^6}} {\cal M}^\prime ({\cal H}/{\mu^6}) \,  \delta {\cal H} , \quad
{\cal M}^\prime  (y)= \dfrac {\text{d}}{\text{d} y} {\cal M} (y)\; . \qquad
\ee
The worldline supersymmetry transformations of the matrix matter fields are
(see Appendix D for their derivation by method described in Appendix C)
\begin{widetext}
 \bea\label{susy=X}
&& \delta_\kappa {\bb X}^i  = \frac {4i}{\sqrt{{\cal M}}}\, \kappa\gamma^i{\mathbf{\Psi}}   + \frac 1 {\mu^6} \,  \frac {{\cal M}^\prime}{{\cal M}} \, \delta_\kappa {\cal H}\;   {\bb X}^i -
\frac 1 {\mu^6} \,  \frac {{\cal M}^\prime}{{\cal M}} \, \Delta_\kappa {\cal K}\,  {\bb P}^i \; , \qquad \\
\label{susy=P} && \delta_\kappa {\bb P}^i =  - \frac {1}{\sqrt{{\cal M}}}\, [\kappa\gamma^{ij}{\mathbf{\Psi}},  {\bb X}^j]
  -\frac 1 {\mu^6} \,  \frac {{\cal M}^\prime}{{\cal M}} \, \delta_\kappa {\cal H}  {\bb P}^i
  +\frac 1 {\mu^6} \,  \frac {{\cal M}^\prime}{{\cal M}} \Delta_\kappa {\cal K}
 \left( \frac 1 {16} [[ {\bb X}^i, {\bb X}^j], {\bb X}^j]-\gamma^i_{pq} \{{\mathbf{\Psi}}_p, {\mathbf{\Psi}}_q\} \right) ,\;  \\
\label{susy=Psi}&& \delta_\kappa {\mathbf{\Psi}}_q=-\frac 1 {2\sqrt{{\cal M}}}\, (\kappa\gamma^{i})_q {\bb P}^i  - \frac {i}{16\sqrt{{\cal M}}}\, (\kappa\gamma^{ij})_q [ {\bb X}^i, {\bb X}^j]
 - \frac {i}  {4\mu^6} \,  \frac {{\cal M}^\prime}{{\cal M}} \, \Delta_\kappa {\cal K}\, [(\gamma^{i}{\mathbf{\Psi}})_q , {\bb X}^i]\; .
\eea
\end{widetext}
Here
\bea \label{kappaH=}
\delta_\kappa {\cal H} =  \frac 1{ \sqrt{{\cal M}}}\, \frac {{\rm tr} \left(\kappa^q{\mathbf{\Psi}}_q\left( [{\bb X}^i, {\bb P}^i] -4i\{{\mathbf{\Psi}}_{q}, {\mathbf{\Psi}}_{q}\}\right) \right) } {1+\frac 1 {\mu^6} \,  \frac {{\cal M}^\prime}{{\cal M}} \, {\frak H} } \; \qquad
\eea
with
\be\label{frakH=} {\frak H}:= \text{tr}\left( {\bb P}^i {\bb P}^i \right)   + {1\over 16}
\text{tr}\left[ {\bb X}^i ,{\bb X}^j \right]^2 + 2\,  \text{tr}\left({\bb X}^i\, \bf{\Psi}\gamma^i {\bf{\Psi}}\right)\;
\ee
is the worldline supersymmetry variation of the relative motion Hamiltonian \eqref{HmM0==} and
\bea \label{kappaK=}
\Delta_\kappa {\cal K} = \frac 1{ 2\sqrt{{\cal M}}}\, \frac { {\rm tr}  \left(4i (\kappa\gamma^i {\bf{\Psi}}) {\bb P}^i + {5\over 2}
(\kappa\gamma^{ij} {\bf{\Psi}})  [{\bb X}^i, {\bb X}^j]  \right)}{1+\frac 1 {\mu^6} \,  \frac {{\cal M}^\prime}{{\cal M}} \, {\frak H} } . \;
\eea
This latter is related to the worldline supersymmetry variation of ${\cal K}= {\rm tr} ({\bb X}^i\,{\bb P}^i)$
by
\be
\label{DkappaK=}
\Delta_\kappa {\cal K}=\delta_\kappa ( {\rm tr} ({\bb X}^i\,{\bb P}^i))+ \frac 1 {2\sqrt{{\cal M}}}i\kappa^q\nu_q
\ee
where
\bea\label{inu=}
i\nu_q&:=& {\rm tr} \left(-4i (\gamma^i {\bf{\Psi}})_q  {\bb P}^i + {1\over 2}
(\gamma^{ij} {\bf{\Psi}})_q  [{\bb X}^i, {\bb X}^j]  \right)\; . \qquad
\eea

In terms of the above blocks the worldline supersymmetry variation of the SU$(N)$ connection 1-form (gauge field)  can be written as (see Appendix D for its derivation)
\begin{widetext}
\bea\label{susy=A}
\delta_\kappa {\bb A} &=&  - \frac 2 {{\cal M}\sqrt{{\cal M}}}\, E^0\,  (\kappa^q{\mathbf{\Psi}}_q) \frac {\left(1- \frac 1 {\mu^6} \, \frac {{\cal M}^\prime}{{\cal M}}{\cal H}\right)}{\left(1+ \frac 1 {\mu^6} \, \frac {{\cal M}^\prime}{{\cal M}}\, {\frak H}\right)} + \frac 1 {\sqrt{2}{\cal M}} \, (E^{1q}-E_q^2)(\gamma^i\kappa)_q \,{\bb X}^i - \qquad  \nonumber \\
& -&  (E^{1q}-E_q^2)\,  \frac 1 {\mu^6} \, \frac {{\cal M}^\prime}{\sqrt{2}{\cal M}^2} \,  \frac  1 {\left(1+ \frac 1 {\mu^6} \, \frac {{\cal M}^\prime}{{\cal M}}\, {\frak H}\right)}  \kappa^p\, {{\mathbf{\Psi}}_{(q}\,{\rm tr}\left(4i (\gamma^i {\mathbf{\Psi}})_{p)} {\bb P}^i +\frac 5 2
(\gamma^{ij} {\bf{\Psi}})_{p)}  [\mathbb{X}^i, {\bb X}^j]  \right) }\,  .
\eea
\end{widetext}

\section{Conclusion and discussion}

Thus, we have found that  the action \eqref{SmD0=} is invariant, besides the manifest spacetime (target superspace type IIA) supersymmetry \eqref{susy=IIA}, also under $16$-parametric local worldline supersymmetry transformations  \eqref{kappa=}, \eqref{susy=X}--\eqref{susy=Psi} and \eqref{susy=A}. Its counterpart in the case of single $p$-branes, local fermionic $\kappa$-symmetry, is considered as an exclusive property of the supersymmetric extended objects of String/M-theory. It guarantees that the ground state of  the dynamical system preserves a part (one-half) of the spacetime supersymmetry.

The form of this worldline supersymmetry depends strongly on the choice of the function ${\cal M}({\cal H}/\mu^6)$ in the action \eqref{SmD0=}. This is restricted by the requirement of non-singularity  ${{\cal M}}\not=0 $ but otherwise is arbitrary \footnote{Similar property is observed in the multiple 0-brane model of \cite{Panda:2003dj} the action of which contains an arbitrary function of matrix matter fields. See \cite{Bandos:2018ntt} for comparison of the properties of this multiple 0-brane model with what one expects for mD$0$ system. }.

The simplest model obtained by setting ${\cal M}=m={\rm const}$ was studied earlier in \cite{Bandos:2018ntt}.
In this case ${\cal M}^\prime =0$ and worldline supersymmetry transformations of the matrix fields \eqref{susy=X}--\eqref{susy=Psi}, \eqref{susy=A} simplify drastically and provides the local supersymmetry generalization of the rigid $d=1$ ${\cal N}=16$ supersymmetry of 10D SU($N$) SYM model reduced to $d=1$.
The local supersymmetry of the action is provided by coupling of this 1d SYM  to the composed worldline supergravity on the worldline induced by the center of energy motion. This is described by 1d graviton 1-form
(einbein) $E^{0}$  and 16 1d gravitini 1-forms $E^{1q}-E_q^2$ constructed from the center of energy variables according to \eqref{E0:=} and \eqref{Eq1=}.

%%%% INSERTION IN V2

Thus the nonlinearity of the previously proposed candidate action with ${\cal M}=m=$const
\cite{Bandos:2018ntt} does not go beyond that of the non-Abelian Yang-Mills.
In contrast the  action \eqref{SmD0=} with a generic function ${\cal M}({\cal H}/\mu^6)$, particularly the one with  \eqref{cM=m+} which can be obtained  by dimensional reduction from 11D mM$0$ action of \cite{Bandos:2012jz}, shows essential nonlinearity beyond the level of SYM one, as it has been  expected
for the multiple D$0$--system. It is impressive that such a nonlinearity can be reached with preserving the local worldline supersymmetry characteristic for mD$0$ system, and that this can be done for essentially arbitrary function ${\cal M}({\cal H}/\mu^6)$.
Also the above mentioned connection with 11D mM$0$ system, the details of which will be  published in a forthcoming paper \cite{Igor+Unai=in-prep}, is another important advantage of the functional \eqref{SmD0=} as a candidate mD$0$ action.

%%%%

The problem of what choice of the function ${\cal M}({\cal H}/\mu^6)$ leads to the true mD$0$-brane action requires additional study. A natural way to make this choice through using T-duality (which was the main argument  for construction of bosonic actions in \cite{Myers:1999ps}) requires as a first step to construct the candidate action for type IIB multiple D1--branes (mD$1$), the problem we are planning to address in the future.  A more detailed study of the properties of the model \eqref{SmD0=} with arbitrary
function ${\cal M}({\cal H}/\mu^6)$, including the solution of its equations of motion and describing its BPS states, can be also useful to single out the true mD$0$-brane action or to clarify why so big set of models possesses the expected properties.

For a moment, an especially interesting in String/M-theoretic perspective looks the model \eqref{SmD0=} with
function ${\cal M}({\cal H}/\mu^6)$ given in  \eqref{cM=m+} because, as we will show in the forthcoming paper \cite{Igor+Unai=in-prep}, this can be obtained by dimensional reduction of the
action for multiple M$0$-brane (multiple M-wave or mM$0$) constructed in \cite{Bandos:2012jz}. However, this argument implies the uniqueness of the action  \cite{Bandos:2012jz} as the one having the properties expected for mM$0$ system. On the other hand, in the light of the found multiplicity of the 10D actions with the properties expected for mD$0$  system, it is tempting to search for possible essentially nonlinear generalizations of
the 11D mM$0$ action of \cite{Bandos:2012jz}.

Also the generalization of the action \eqref{SmD0=} for the case of multiple D$p$-brane system with $1 < p\leq 9$ and for the case of curved target IIA supergravity superspace are intriguing and important problems.

\smallskip

\noindent\textbf{Acknowledgements}:
 The work by IB  was partially supported by Spanish MINECO and FEDER (ERDF EU) under  grant PGC2018-095205-B-I00 and by the Basque Government Grant IT1628-22.

%%%%APPENDIX
\appendix

\section{10D spinor moving frame variables}
%and the action for single D$0$-brane}

The multiple D$0$-brane action, presented in the main text, is presently known only in its spinor moving frame formulation involving the auxiliary variables which we are going to describe in some details.

The Spin$(1,9)/$Spin$(9)$ spinor moving frame variables and their moving frame vector companions appropriate to the description of D0 brane and multiple D0 (mD0) systems are elements of, respectively, 16$\times$16 and 10$\times$10 matrices \eqref{harmV=10} and \eqref{harmU=10} (see \cite{Bandos:2000tg} and \cite{Bandos:2018ntt})
\begin{eqnarray}\label{harmV=10D0}
v_\alpha{}^q \in \text{Spin}(1,9)  \qquad {\rm and} \qquad (u_\mu^0, u_\mu^i) \in \text{SO}(1,9)\; .
\end{eqnarray}
Here $i = 1,...,9$ and $q= 1,...,16$ are vector and spinor indices of SO$(9)$ group while
$\mu,\nu = 0,1,...,9$  and $\alpha, \beta = 1,2,...,16$ are  $10$-vector and 10D Majorana-Weyl spinor indices.

The condition that moving frame variables form the SO$(1,9)$ valued matrix implies \eqref{u0u0=1}
% that \be u_\mu^0 u^{\mu 0}=1 , \qquad u_\mu^0 u^{\mu i}=0\; , \qquad u_\mu^iu^{\mu j}=-\delta^{ij}\; , \ee
and
\be
u_\mu^0u_\nu^0- u_\mu^iu_\nu^i= \eta_{\mu\nu} = {\rm{diag}} (1,-1,\ldots ,-1)\; .
\ee

The spinor moving frame variables obey the constraints
\be u^{(\nu)}_\mu \sigma^\mu_{\alpha\beta}= v_{\alpha}^q \sigma^{(\nu)}_{qp}v_{\beta}^p\; , \qquad
u^{(\nu)}_\mu \tilde{\sigma}{}_{(\nu)}^{qp}= v_{\alpha}^q \tilde{\sigma}{}_{\mu}^{\alpha\beta}v_{\beta}^p\qquad
\label{eq:rel}
\ee
which express the SO$(1,9)$ Lorentz invariance of the 10D generalization of the relativistic Pauli matrices
$\sigma^\mu_{\alpha\beta}= \sigma^\mu_{\beta\alpha}$ and $\tilde{\sigma}{}_{\mu}^{\alpha\beta}=\tilde{\sigma}{}_{\mu}^{\beta\alpha}$,
\bea
\sigma^{(\mu}_{\alpha\gamma}\tilde{\sigma}{}^{\nu)\gamma\beta}:= \frac 1 2
\left(\sigma^{\mu}\tilde{\sigma}{}^{\nu}+\sigma^{\nu}\tilde{\sigma}{}^{\mu}\right)_{\alpha}{}^{\beta} = \eta^{\mu\nu}\delta_{\alpha}{}^{\beta}\; , \qquad
\eea
and also makes the spinor frame matrix to describe double covering of the Lorentz group element represented by the moving frame matrix
(see \cite{Bandos:1990ji,Delduc:1991ir,Galperin:1991gk}). Roughly speaking this statement can be formulated by saying that spinor frame variables
(also called Lorentz harmonics \cite{Bandos:1990ji,Delduc:1991ir,Galperin:1991gk}) are square roots of the moving frame variables (also called vector harmonics \cite{Sokatchev:1985tc}).

Choosing the SO$(9)$ invariant representation
\be \sigma^{(\mu)}_{qp}=(\delta_{qp}, \gamma^i_{qp})=\tilde{\sigma}{}_{(\mu)}^{qp}\; , \qquad \ee where $\gamma^i_{qp} = \gamma^i_{pq}$ are $d=9$ gamma matrices,
\be
\gamma^i_{qp}=\gamma^i_{pq}\; , \qquad \gamma^{(i}\gamma^{j)}=\delta^{ij}{\bb I}_{16\times 16}~,
\ee
we find that Eqs. \eqref{eq:rel} acquire the form of \eqref{u0s=vv} and
\be
v_{\alpha}^q \tilde{\sigma}{}_{\mu}^{\alpha\beta}v_{\beta}^p= u_\mu^{{0}} \delta_{qp}+u_\mu^{{i}} \gamma^i_{qp}\; .
\ee
Similarly,  we find
\bea
\label{u0ts=vv}
 u_\mu^0 \tilde{\sigma}^{\mu\alpha\beta}=v_q{}^\alpha v_q{}^\beta \; , \qquad
u_\mu^i \tilde{\sigma}^{\mu \alpha\beta}=- v_q{}^\alpha  \gamma^i_{qp} v_p{}^\beta\;~. \qquad
\eea
Notice that
\begin{eqnarray}\label{vs=v-1}
v_q{}^\beta= v_\alpha^q \tilde{\sigma}^{\mu\alpha\beta} u_\mu^0 \; , \quad{\rm obeying} \quad   u_\mu^0 {\sigma}^{{\mu}}_{\alpha\beta}v_q{}^\beta  = v_\alpha{}^q ~ ,
\end{eqnarray}
is the inverse spinor moving frame matrix $v_\alpha{}^q\in$ Spin$(1,9)$:
\begin{eqnarray}\label{harmV-1=10D0}
v_q{}^\alpha v_\alpha{}^p=\delta_q{}^p \qquad \Leftrightarrow \qquad
 v_\alpha{}^q v_q{}^\beta= \delta_\alpha{}^\beta
\; . \qquad
\end{eqnarray}

The derivatives of the moving frame and of the spinor moving frame variables are expressed in terms of Cartan forms
\begin{equation}
\begin{array}{ccc}
\Omega^i = u_\mu^0 \text{d}u^{\mu i}~,&~&\Omega^{ij} = u_\mu^i \text{d}u^{\mu j}\qquad
\end{array}
\label{Omi=}
\end{equation}
by
\begin{equation}
\begin{array}{ccc}
\text{D}u^0_\mu := \text{d}u_\mu^0 =u_\mu^i \Omega^i~,&~&\text{D}u^i_\mu := \text{d}u_\mu^i + u_\mu^j\Omega^{ji} =u_\mu^0 \Omega^i~
\end{array}
\label{Du0=}
\end{equation}
and
\begin{eqnarray}\label{Dv=vOm}
&&\text{D}v_\alpha{}^q:= \text{d}v_\alpha{}^q+ \frac 1 4 \Omega^{ij} v_\alpha{}^p\gamma^{ij}_{pq}
= \frac 1 2 \gamma^i_{qp} v_\alpha{}^p\Omega^i \; \qquad
\\
&& \Rightarrow \;
\text{D}v_q{}^\alpha := \text{d}v_q{}^\alpha-  \frac 1 4 \Omega^{ij} \gamma^{ij}_{qp}v_p^\alpha
= -\frac 1 2 v_p^\alpha\gamma^i_{pq} \Omega^i\; . \qquad
\label{Dv=vOm}
\end{eqnarray}
Taking exterior derivatives of Eqs. \eqref{Du0=} (see Appendix \ref{sec:DF} for definitions) we can find the Maurer-Cartan equations
\bea\label{MC=Eq}
\text{D}\Omega^i=\text{d} \Omega^i+ \Omega^j\wedge \Omega^{ji} = 0 \; , \qquad \nonumber  \\ \text{d}\Omega^{ij}+\Omega^{ik}\wedge \Omega^{kj} =-\Omega^i\wedge \Omega^j\; . \quad
\eea

\section{Single D0-brane in spinor moving frame formulation and its $\kappa$-symmetry}
\label{sec:kappa}
The action of the moving frame formulation of the 10D D0-brane  in flat type IIA superspace, which also appears as a part of the multiple D$0$-brane action \eqref{SmD0=} describing the center of mass dynamics of this system,  reads \cite{Bandos:2000tg}
\begin{equation}\label{eq:L_D0}
S_{\text{D}0} = \int\limits_{\mathcal{W}^1}{\cal L}_{\text{D}0}= m\int_{\mathcal{W}^1} E^0 -im \int\limits_{\mathcal{W}^1} \left(\text{d} \theta^{1 \alpha}\theta_{\alpha}^{2} - \theta^{1 \alpha}\text{d}\theta_{\alpha}^{2} \right)~.
\end{equation}
Here $\text{d} =  \text{d}\tau\partial/\partial \tau =:\text{d}\tau \partial_\tau$, $\tau$ is proper time variable parametrizing the D0-brane  worldline
$\mathcal{W}^1$  defined as a line  in target $D=10$ type IIA superspace $\Sigma^{(10|32)}$ with $10$ bosonic and $16+16=32$ fermionic coordinates
\be\label{ZM} Z^M=( x^\mu, \theta^{1 \alpha}, \theta_{\alpha}^2) \qquad  \ee
by corresponding coordinate functions
\bea\label{ZMt} && Z^M(\tau)=( x^\mu(\tau), \theta^{1 \alpha}(\tau), \theta_{\alpha}^2 (\tau))\; , \qquad \\ \label{cW1}
 \mathcal{W}^1 \in \Sigma^{(10|32)} &: &\qquad Z^M=Z^M(\tau)\; . \qquad
\eea
The constant $m$ entering both terms of \eqref{eq:L_D0} is the mass of D$0$-brane and $E^0$ is the contraction
\begin{equation}\label{E0=}
E^0 = \Pi^\mu u^0_\mu\qquad
\end{equation}
of the pull-back to the worldline
of the 10D Volkov-Akulov $1$-form
\begin{equation}
\Pi^\mu = \text{d}x^\mu -i\text{d}\theta^1\sigma^\mu \theta^1 -i\text{d}\theta^2\tilde{\sigma}^\mu \theta^2 \qquad
\end{equation}
with the vector field $u_\mu^0= u_\mu^0(\tau)$.
The pull-back of a differential form on target superspace is obtained by substituting the coordinate functions for coordinates; so that
Eq. \eqref{E0=} actually includes
 \begin{equation}
\Pi^\mu =\text{d}\tau \Pi_\tau^\mu = \text{d}x^\mu(\tau) -i\text{d}\theta^1(\tau)\sigma^\mu \theta^1(\tau) -i\text{d}\theta^2(\tau)\tilde{\sigma}^\mu \theta^2(\tau)\; .
\end{equation}
Notice that,  to simplify notation, below and below, as well as in the main text, we use the same symbols for
the differential forms on the target superspace and their pull-backs to the worldline $\mathcal{W}^1$.
The same applies to the superspace coordinates \eqref{ZM} and the
coordinate functions \eqref{ZMt}. Particularly, in the second term of \eqref{eq:L_D0}  $\theta^{1 \alpha}$ and $\theta_{\alpha}^2$ denote  $\theta^{1 \alpha}(\tau)$ and $\theta_{\alpha}^2(\tau)$.

A very important property of the action  (\ref{eq:L_D0}) is that, besides manifest $D=10$ $\mathcal{N}=2$ spacetime supersymmetry, it is also invariant under the following local fermionic $\kappa$-symmetry transformations
\begin{equation}
\begin{array}{ccl}
\delta_\kappa \theta^{1 \alpha}=   \kappa^q v_q^\alpha ~,~~~~~ \delta_\kappa \theta_{\alpha}^{2}= -  \kappa^q v_\alpha{}^q~,
\\ \delta_\kappa x^\mu =i\delta_\kappa\theta^1 \sigma^\mu \theta^1+i\delta_\kappa\theta^2 \tilde{\sigma}{}^\mu \theta^2~,\\
%~\\
\delta_\kappa v_\alpha{}^q=0 \qquad \Rightarrow \qquad \delta_\kappa u^i_\mu = \delta_\kappa u^0_\mu = 0~~  ,
%~\\
\end{array}
\label{eq:kappaD0=}
\end{equation}
where $\kappa^q =\kappa^q (\tau)$ with $q=1,...,16$ are arbitrary  fermionic functions.

To prove the $\kappa$-invariance of the single D$0$-brane action and also the invariance of multiple D$0$-brane action under its generalization, the worldline supersymmetry, we have used the formalism of generalized Lie derivatives based on formal exterior derivatives of differential forms which we are going to describe in the next Appendix C.

\section{Differential forms and variations}
\label{sec:DF}

%In this section we review some basic concepts of the differential forms and its exterior derivative formalism.

Let $\Xi_q$ be differential $q$-form in a superspace with coordinates $Z^{M}$,
\begin{equation}
\Xi_q = \dfrac{1}{q!}~\text{d}Z^{M_q} \wedge ... \wedge \text{d}Z^{M_1} \Xi_{M_1 ... M_q}(Z)
\end{equation}
where $\wedge$ is the exterior product of the differential forms. In the simplest case of basic 1-forms given by differentials of the superspace coordinates,
\begin{equation}\label{dZwdZ}
\text{d}Z^M \wedge \text{d}Z^N = - (-1)^{\epsilon(M)\epsilon(N)}~\text{d}Z^N \wedge \text{d}Z^M~,
\end{equation}
where $\epsilon(M) \equiv \epsilon(Z^M)$ is the so-called Grassmann parity of $Z^M$ defined by
\begin{equation}
\begin{array}{ccccc}
\epsilon(x^\mu) = 0~,&~&\epsilon(\theta^{1 \alpha}) = 1~,&~&\epsilon(\theta^{2}_\alpha) = 1~
\end{array}
\end{equation}
in the case of $D=10$ type IIA superspace with  coordinates $Z^M=(x^\mu, \theta^{ \alpha 1}, \theta^2_{\alpha})$.
For any bosonic $p$- and $q$-forms
\begin{equation}
\Xi_q \wedge \Upsilon_p = (-1)^{qp} \Upsilon_p \wedge \Xi_q~,
\end{equation}
in particular,
\be
\text{d}x^\mu\wedge \text{d}x^\nu= - \text{d}x^\nu\wedge \text{d}x^\mu\; .
\ee

In the case of the forms which can be also fermionic
\begin{equation}
\Xi_q \wedge \Upsilon_p = (-1)^{qp+  \epsilon(\Xi_q)\epsilon(\Upsilon_p)} \Upsilon_p \wedge \Xi_q~\; .
\end{equation}
In particular, \eqref{dZwdZ} implies that all products of the supercoordinate differentials are antisymmetric but
\bea
\text{d}\theta^{1 \alpha} \wedge \text{d}\theta^{\beta 1} = \text{d}\theta^{\beta 1} \wedge \text{d}\theta^{1 \alpha}, \quad \text{d}\theta_{\alpha}^{2} \wedge \text{d}\theta_{\beta}^{2} = \text{d}\theta_{\beta}^{2} \wedge \text{d}\theta_{\alpha}^{2}~, \nonumber \\ \text{d}\theta^{1 \alpha} \wedge \text{d}\theta_{\beta}^{2} = \text{d}\theta_{\beta}^{2} \wedge \text{d}\theta^{1 \alpha}~. \qquad \nonumber
\eea

The exterior derivative of the differential forms, which maps $q$-forms into ($q+1$)-forms, is defined by
\bea
& \text{d}\Xi_q = \dfrac{1}{q!}~\text{d}Z^{M_q} \wedge ... \wedge \text{d}Z^{M_1} \wedge \text{d}Z^{M_0} \partial_{M_0} \Xi_{M_1...M_q}(Z) \equiv \qquad  \nonumber \\
&\equiv  \dfrac{1}{(q+1)!}~\text{d}Z^{M_{q+1}} \wedge ... \wedge \text{d}Z^{M_1} \times \qquad \nonumber \\ & \times (q+1) \partial_{\left[ M_1 \right.} \Xi_{\left. M_1...M_{q+1} \right\rbrace}(Z)~,\nonumber
\eea
where $\partial_N = \frac{\partial}{\partial Z^N}$ and $\left[...\right\rbrace$ denotes graded antisymmetrization over the enclosed indices, in particular
\be
\Xi_{\left[ MN \right\rbrace}= \frac 1 2 \left(\Xi_{MN}-(-)^{\epsilon(M)\epsilon(N) } \Xi_{NM}\right) \; .
\ee
The exterior derivative operator d  obeys the nilpotency condition and the (generalized) Leibniz rule
\begin{equation}
\begin{array}{cccc}
\text{d}\text{d} = 0 ~,&~&\text{d} (\Xi_q \wedge \Xi_p) = \Xi_q \wedge \text{d}\Xi_p + (-1)^p~\text{d}\Xi_q \wedge \Xi_p~.
\end{array}
\end{equation}

The variation of differential forms under generic  transformations of coordinates can be calculated using the so-called Lie derivative formula,
\begin{equation}
\delta \Xi_q = i_{\delta}\left(\text{d}\Xi_q \right) + \text{d}\left(i_\delta \Xi_q \right)~,
\end{equation}
where $i_\delta$ is the contraction with variation symbol defined by
\begin{equation}
i_\delta \Xi_q = \dfrac{1}{(q-1)!}~\text{d}Z^{M_q} \wedge ... \wedge\text{d}Z^{M_2} \delta Z^{M_1}\Xi_{M_1...M_q}(Z)~.
\end{equation}
Notice that this implies
\be
i_\delta \text{d}{Z^{M}}=  \delta Z^{M}\; . \qquad
\ee

The contraction $i_\delta$  maps differential $q$-forms into  $(q-1)$-forms and obeys its own counterpart of the Leibnitz rule:
\begin{equation}
i_\delta (\Xi_q \wedge \Xi_p) = \Xi_q \wedge i_\delta \Xi_p + (-1)^p~i_\delta \Xi_q \wedge \Xi_p~.
\end{equation}

The variation of the Lagrangian D-form $\mathcal{L}$ of a D-dimensional field theory can be calculated using the Lie derivative formula
with formal exterior derivative\footnote{Here `formal' means that we do not use the formula
${\rm{d}}={\rm{d}} \tau\partial_\tau$ in our case or its D-dimensional generalization in the case of D-dimensional field theory; if we did, this would clearly imply vanishing of any 2-form in our case or any (D+1)-form in D-dimensional space.  In other words, the procedure implies formal extension  of all the differential forms from the worldline to target superspace or, better to say, to some its extension, some supergroup manifold which also includes the coordinates corresponding to  spinor moving frame variables (called Lorentz harmonic superspace in
\cite{Bandos:1990ji}; see also \cite{Sokatchev:1985tc}). The differentials of these latter are expressed in terms of Cartan forms \eqref{Omi=}. } \begin{equation}
\delta \mathcal{L} = i_\delta (\text{d} \mathcal{L}) + \text{d}(i_\delta \mathcal{L})~.
\label{eq:LieDer_L}
\end{equation}
The total derivative term $\text{d}(i_\delta \mathcal{L})$  is not essential when we derive the equations of motion and can be conventionally omitted if one does not study effects of boundary contributions.

In the models with manifest gauge symmetry
it is more convenient to define the variations of differential forms  given by covariant Lie derivative
\begin{equation}
\delta \Xi^{{\cal A}}_q = i_\delta (\text{D} \Xi^{{\cal A}}_q) + \text{D}(i_\delta \Xi^{{\cal A}}_q)~,
\end{equation}
where $\text{D}$ is covariant derivative including the connection of the gauge symmetry group  and ${\cal A}$ is an index (or multi-index including the index) of a representation of the gauge group carried by the differential $q$-form. Clearly for the Lagrangian D-form, which is invariant under the gauge symmetry, D${\cal L}= \text{d}{\cal L}$ and the covariant Lie derivative prescription coincides with the standard Lie derivative one \eqref{eq:LieDer_L}.

As a warm-up exercise  let us apply this method to vary the Lagrangian 1-form  of the action \eqref{eq:L_D0} of single  D0-brane in flat 10D type IIA superspace \cite{Bandos:2000tg}:
\begin{equation*}
\mathcal{L}_{\text{D}0} = mE^0 - im (\text{d}\theta^1 \theta^2-\theta^1 \text{d}\theta^2)\qquad
\end{equation*}
with constant $m$.

The formal exterior derivative  of  $E^0 = \Pi^\mu u^0_\mu$ in the first term of the Lagrangian form  is given by
\begin{equation}
\text{d}E^0 = E^i \wedge \Omega^i - i\left(E^{1q}\wedge E^{1q} + E^{2}_q \wedge E^{2}_q \right)~,
\end{equation}
where
\begin{equation}\label{Ei=}
E^i = \Pi^\mu u^i_\mu\; , \qquad E^{1q}= \text{d}\theta^{1\alpha}\, v_\alpha{}^q\; ,   \qquad E_q^2= \text{d}\theta_\alpha^2 v_q{}^{\alpha}\; . \qquad
\end{equation}
To find that we have used
\begin{equation}
\text{d}\Pi^\mu = -i\text{d}\theta^1\sigma^\mu \wedge \text{d}\theta^1 -i\text{d}\theta^2\tilde{\sigma}^\mu \wedge \text{d}\theta^2~
\end{equation}
as well as Eqs. \eqref{u0s=vv} and  \eqref{Du0=}.

 The derivative of the second,  Wess-Zumino term of the D$0$-brane action is
\be
-2im \text{d}\theta^{1\alpha} \wedge \text{d}\theta_\alpha^2= -2im E^{1q}\wedge E_{q}^2\; .
\ee

Now after an elementary algebra  we find that the formal exterior derivative of the Lagrangian form of single D$0$-brane can be written as
\begin{equation}
\text{d}{\cal L}_{\text{D}0}= mE^i\wedge \Omega^i -im (E^{1q}+E^2_{q})\wedge (E^{1q}+E^2_{q})~,
\end{equation}
where $\Omega^i$ is the covariant Cartan form defined in \eqref{Omi=}.

Then, using the Lie derivative formula (\ref{eq:LieDer_L}), we find
\bea\label{vcLD0=}
\delta \mathcal{L}_{\text{D}0} = m\left(E^i i_\delta \Omega^i - i_\delta E^i \Omega^i \right) - \qquad \nonumber \\ - 2im\left(E^{1q} + E^2_q \right)\left(i_\delta E^{1q} + i_\delta E^2_q \right)~,
\eea
where $i_\delta \Omega^i$ defines essential variation of the spinor frame variable by
%\begin{eqnarray}\label{Dv=vOm}
$\delta v_\alpha{}^q =i_\delta \text{D}v_\alpha{}^q
=  \frac 1 2\gamma^i_{qp} v_\alpha{}^p i_\delta \Omega^i$.
% \;. \qquad
%\end{eqnarray}
This equation can be obtained from  the $i_\delta$ contraction of \eqref{Dv=vOm} by setting  $ i_\delta\Omega^{ij}=0$.

To conclude, let us note that in this formalism the local fermionic $\kappa$-symmetry transformations $\delta_\kappa$ \eqref{eq:kappaD0=} leaving invariant the D$0$-brane action  (\ref{eq:L_D0}) can be described by
($i_\kappa {\rm d}:=\delta_\kappa$)
\begin{widetext}
\begin{equation}
\begin{array}{ccl}
 i_\kappa \Pi^{\mu}=\delta_\kappa x^\mu -i\delta_\kappa\theta^1 \sigma^\mu \theta^1 - i\delta_\kappa\theta^2 \tilde{\sigma}{}^\mu \theta^2=0\qquad \Rightarrow \qquad  i_\kappa E^0=0~, \qquad  i_\kappa E^i=0 ~,\\
%~\\
i_\kappa\Omega^i=0~,
\qquad i_\kappa\Omega^{ij}=0~,\\
%~\\
i_\kappa E^{1q}= - i_\kappa E^{2}_{q}= \kappa^q \qquad \Rightarrow \qquad i_\kappa (E^{1q}+ E^{2}_{q})=0 \; .
\end{array}
\label{eq:kappaD0}
\end{equation}
Indeed substituting the above $i_\kappa$ for $i_\delta$ in \eqref{vcLD0=}, we find $\delta_\kappa {\cal L}_{\text{D}0}=0$.

\section{Multiple D$0$-brane action and its worldline supersymmetry}

In this Appendix  we present some details of the derivation of the  worldline supersymmetry leaving invariant the candidate mD$0$ action \eqref{SmD0=}.

\subsection{Formal exterior derivative of the  Lagrangian form of the mD$0$ action}

The  first stage is to calculate the formal exterior derivative of the Lagrangian form of the action \eqref{SmD0=}, this is to say of 1-form
\begin{equation}\label{LmD0==}
\begin{array}{c}
\begin{split}
\mathcal{L}_{\text{mD}0} &= mE^0 - im (\text{d}\theta^1 \theta^2-\theta^1 \text{d}\theta^2)~+\\
%&~\\
&+ \frac 1 {\mu^6}  \left[  \text{tr}\left({\bb P}^i \text{D} {\bb X}^i + 4i {\mathbf{\Psi}}_q \text{D}
{\mathbf{\Psi}}_q  \right) +  \frac 2 {\cal M} {E}^{0} {\cal H}-   \frac {\text{d} {\cal M}}{ {\cal M} } \text{tr} ({\bb P}^i{\bb X}^i)~+\right.\\
%&~\\
&\left.{}\qquad +\frac 1 {\sqrt{2{\cal M}}}({E}{}^{q1}-{E}{}^{2}_{q}) \text{tr} \left(-4i (\gamma^i {\mathbf{\Psi}})_q  {\bb P}^i + {1\over 2}
(\gamma^{ij} {\mathbf{\Psi}})_q  [{\bb X}^i, {\bb X}^j]  \right)\right]~,
\end{split}
\end{array}
\end{equation}
where ${\cal H}$ is given  in Eq. \eqref{HmM0==}.
The covariant derivatives D of the bosonic and fermionic Hermitian traceless $N \times N$ matrix fields are defined in \eqref{DXi=} and \eqref{DPsi:=}
 with the use of $1$d gauge field 1-form $\mathbb{A}= \text{d}\tau \mathbb{A}_\tau $ and Cartan forms (\ref{Omi=}),
so that, when calculating the  exterior derivative of \eqref{LmD0==}, we have to use the Ricci identities
\bea
\text{DD}{\bb X}^i= \Omega^i\wedge \Omega^j \, {\bb X}^j + [{\bb F}, {\bb X}^i]~, \qquad
%\\
\text{DD}{\mathbf{\Psi}}_q = \frac 1 4 \, \Omega^i\wedge \Omega^j\, (\gamma^{ij}{\mathbf{\Psi}})_q  + [{\bb F}, {\mathbf{\Psi}}_q]~. \label{Ricci}
\eea
Here  $\mathbb{F} = \text{d}\mathbb{A} - \mathbb{A} \wedge \mathbb{A}$ is the formal 2-form field strength of the 1d gauge field $\mathbb{A}$ (which is calculated without using $\mathbb{A}=\text{d}\tau \mathbb{A}_\tau $, with the aim to apply it in the Lie derivative formula for variation of the Lagrangian 1-form). Eqs. \eqref{Ricci} are obtained using the Maurer-Cartan equations \eqref{MC=Eq}.

After some algebra, the exterior derivative of the multiple D$0$-branes Lagrangian form \eqref{LmD0==} can be found to be
\begin{equation}\label{dLmD0==}
\begin{array}{c}
\begin{split}
\mu^6\text{d}{\cal L}_{\text{mD}0}&= \mu^6 m E^i\wedge \Omega^i -i\mu^6 m (E^{1q}+E^2_{q})\wedge (E^{1q}+E^2_{q})
+ \Omega^i\wedge \Omega^j \, {\rm tr} ({\bb P}^i {\bb X}^j+ i\mathbf{\Psi}\gamma^{ij}\mathbf{\Psi})~+\\
%&~\\
&+{\rm tr} \left({\bb F} \left(\left[{\bb X}^i,{\bb P}^i\right]- 4i \{\mathbf{\Psi}_q,\mathbf{\Psi}_q\}\right)\right)  -  {\rm tr} (\text{D}{\bb P}^i \wedge \text{D}{\bb X}^i) -4i {\rm tr} (\text{D}{\mathbf{\Psi}}_q \wedge \text{D}{\mathbf{\Psi}}_q)~+\\
%&~\\
&+ \frac 2 {{\cal M}} E^i\wedge \Omega^i {\cal H} -i \frac 2 {{\cal M}}  ( E^{1q}\wedge E^{1q}+ E^2_{q}\wedge E^2_{q}) {\cal H}  -  \frac 1 {2\sqrt{2{\cal M}}}  (E^{1q}+E_q^2) \gamma^i_{qp}i\nu_p \wedge \Omega^i~+\\
%&~\\
& + \frac 2 {{\cal M}} \left(1- \frac 1 {\mu^6}  \frac {{\cal M}^\prime}{{\cal M}}{\cal H} \right) E^0\wedge \text{d}{\cal H}  + \frac 1 {\sqrt{2{\cal M}}} (E^{1q}-E_q^2)\wedge i\text{D}\nu_q + \frac 1 {\mu^6}  \frac {{\cal M}^\prime}{{\cal M}}  \text{d}{\cal K} \wedge \text{d}{\cal H}~+\\
%&~\\
&+ \frac 1 {\mu^6} \, \frac 1 {2\sqrt{2{\cal M}}} \frac {{\cal M}^\prime}{{\cal M}} (E^{1q}-E_q^2) i\nu_q \wedge \text{d}{\cal H}~,
\end{split}
\end{array}
\end{equation}
where ${\cal K}:= {\rm tr} ({\bb X}^i{\bb P}^i)$, $\nu_q$ is defined in \eqref{inu=}
and ${\cal H}$ is the relative motion Hamiltonian \eqref{HmM0==}.
The derivatives of these `blocks', which also enter \eqref{dLmD0==},  read
\begin{eqnarray}
\label{dHmM0==}  \text{d}{\cal H} &=&  \text{tr}\left( {\bb P}^i \text{D} {\bb P}^i+ \frac 1 {16} \text{D} {\bb X}^i [[ {\bb X}^i, {\bb X}^j], {\bb X}^j] - \text{D} {\bb X}^i \gamma^i_{pq} \{{\mathbf{\Psi}}_p, {\mathbf{\Psi}}_q\}    - 2\, \text{D}{\mathbf{\Psi}}_q  [(\gamma^i {\mathbf{\Psi}})_q , {\bb X}^i]\, \right) , \qquad
\\
\text{d}{\cal K}&=& {\rm tr} (\text{D}{\bb X}^i\,{\bb P}^i+{\bb X}^i\text{D}{\bb P}^i)\; , \qquad \\
  i\text{D}\nu_q&=& {\rm tr} \left(-4i (\gamma^i \mathbf{\Psi})_q  \text{D}{\bb P}^i -4i (\gamma^i \text{D}\mathbf{\Psi})_q  {\bb P}^i -
\text{D}{\bb X}^i[(\gamma^{ij} \mathbf{\Psi})_q  , {\bb X}^j] + {1\over 2}
(\gamma^{ij} \text{D}\mathbf{\Psi})_q  [{\bb X}^i, {\bb X}^j]  \right) .
\eea

%\subsection{Important variations for studying the worldline supersymmetry}
%In this section we collect the most important variations which are basic in order to get the local worldline supersymmetry of the mD0 system.

\subsection{Worldline supersymmetry ($\kappa$-symmetry) transformations of the center of energy variables}
%and the variation of the Lagrangian form}

The previous experience with lower-dimensional counterparts of the mD$0$ system~\cite{Bandos:2021vrq}  suggests  to assume that the worldline supersymmetry acts on the center of energy variables of the mD$0$ system (i.e. on the superspace coordinate functions and spinor frame variables)
as the $\kappa$-symmetry of the single D$0$-brane action (see sec.~\ref{sec:kappa}) acts on their single-brane counterparts. Namely, we set
\footnote{The re-scaling of fermionic function ${\kappa^q}\mapsto \frac {\kappa^q} {\sqrt{2}}$ is performed to simplify the worldline supersymmetry  transformation rules of the matrix fields.}
\begin{equation}\label{ikPi=0}
\begin{array}{ccl}
 i_\kappa \Pi^{\mu}=0 & \Rightarrow & \quad i_\kappa E^0=0\; , \qquad  i_\kappa E^i=0\; , \qquad \\
%~\\
i_\kappa\Omega^i=0~,
\qquad i_\kappa\Omega^{ij}=0& \Rightarrow &\quad \delta_\kappa u^0_\mu = 0\; , \qquad \delta_\kappa u^i_\mu = 0 \; , \qquad \delta_\kappa v_\alpha{}^q=0\; ,
\end{array}
\end{equation}
and
\begin{equation}\label{ikEq1=}
i_\kappa E^{1q}= - i_\kappa E^{2}_{q}= \frac {\kappa^q} {\sqrt{2}}\qquad \Rightarrow \qquad \delta_\kappa \theta^{1 \alpha}=  \frac {\kappa^q} {\sqrt{2}}v_q^\alpha ~, \qquad \delta_\kappa \theta_{\alpha}^{2}= - \frac {\kappa^q} {\sqrt{2}}v_\alpha{}^q~.
\end{equation}
These expressions are equivalent to \eqref{kappa=}, but they are more convenient to use in our method of calculation of the variation of Lagrangian form.

Then, using the Lie derivative formula  (\ref{eq:LieDer_L}) with \eqref{ikPi=0}, \eqref{ikEq1=} and furthermore identifying in it
\begin{equation}
i_\kappa \text{D}=\delta_\kappa~, \qquad i_\kappa {\bb F}=\delta_\kappa {\bb A}~, \qquad i_\kappa {\bb A}=0~,
\end{equation}
we find that, modulo total derivative, the variation $\delta_\kappa$ of the Lagrangian form ${\cal L}_{\text{mD}0}$
reduces to
\begin{equation}
\begin{array}{c}
\begin{split}
\mu^6\delta_\kappa {\cal L}_{\text{mD}0}&=  \text{tr}\left(\delta_\kappa \mathbb{A} \left(\left[{\bb X}^i,{\bb P}^i\right]- 4i \{\mathbf{\Psi}_q,\mathbf{\Psi}_q\}\right)\right) + {\rm tr} \left(\delta_\kappa{\bb P}^i \text{D}{\bb X}^i - \text{D}{\bb P}^i \delta_\kappa {\bb X}^i-8i  \text{D}{\mathbf{\Psi}}_q \delta_\kappa {\mathbf{\Psi}}_q\right) ~-\\
%&~\\
&- i \frac {2\sqrt{2}} {{\cal M}}  ( E^{1q}- E^2_{q})\kappa^{q}{\cal H} + \frac 2 {{\cal M}} \left(1- \frac {{\cal H}} {\mu^6} \, \frac {{\cal M}^\prime}{{\cal M}}\right) E^0\delta_\kappa {\cal H}+  \frac 1 {\mu^6} \,  \frac {{\cal M}^\prime}{{\cal M}}  \text{d}{\cal K} \delta_\kappa {\cal H}~+\\
%&~\\
&+ \frac 1 {\mu^6} \, \frac 1 {2\sqrt{2{\cal M}}} \frac {{\cal M}^\prime}{{\cal M}} \,  (E^{1q}-E_q^2) i\nu_q \delta_\kappa {\cal H} - \frac 1 {\mu^6} \,  \frac {{\cal M}^\prime}{{\cal M}} \,  \delta_\kappa{\cal K}\,  \text{d}{\cal H}~-\\
%&~\\
&-\frac 1 {\mu^6} \, \frac 1 {2\sqrt{{\cal M}}} \frac {{\cal M}^\prime}{{\cal M}} \, \kappa^q i\nu_q\,  \text{d}{\cal H} - \frac 1 {\sqrt{{\cal M}}} \kappa^q i\text{D}\nu_q
 + \frac 1 {\sqrt{2{\cal M}}} (E^{1q}-E_q^2)\, i\delta_\kappa\nu_q~.
\end{split}
\end{array}
\label{susy=cLmD0}
\end{equation}
The worldline supersymmetry transformation rules of the matrix fields can be found by requiring this variation to vanish.
As this calculation is a bit subtle, we present below some details.

\subsection{Worldline supersymmetry transformations of the matrix matter fields }

To find the  supersymmetry transformation leaving invariant $S_{\text{mD}0}= \int {\cal L}_{\text{mD}0}$, i.e. obeying $\delta_\kappa {\cal L}_{\text{mD}0}=0$ (modulo total derivative),   we have to set equal to zero the coefficients for all the independent 1-forms in \eqref{susy=cLmD0}. Requiring to vanish the terms proportional to D${\bb P}^i$, D${\bb X}^i$ and  D${\mathbf{\Psi}}_q$, we find the  set of {\it equations} for the worldline supersymmetry transformations of the matrix `matter' fields of the form of relations
 \eqref{susy=X}, \eqref{susy=P} and \eqref{susy=Psi}.

We stress that at this stage these are equations because their right hand sides contain $\Delta_\kappa {\cal K}$ from \eqref{DkappaK=} and $\delta_\kappa {\cal H}$ which in their turn are expressed in terms of
$\delta_\kappa {\bb X}^i$, $\delta_\kappa {\bb P}^i$ and $\delta_\kappa {\mathbf{\Psi}}_q$.

To solve these equations it is convenient to calculate formally the variations of composite quantities
$\delta_\kappa {\cal H}$ and $\Delta_\kappa {\cal K}$ with \eqref{susy=X}-\eqref{susy=Psi}.
On this way we find the following equations
\bea
\Delta_\kappa {\cal K} &=& \frac 1{ 2\sqrt{{\cal M}}}{\rm tr} \left(4i (\kappa\gamma^i \mathbf{\Psi}) {\bb P}^i + {5\over 2}
(\kappa\gamma^{ij}\mathbf{\Psi}) [{\bb X}^i, {\bb X}^j]  \right) - \frac 1 {\mu^6} \,  \frac {{\cal M}^\prime}{{\cal M}} \,\Delta_\kappa {\cal K}\, {\frak H}~, \qquad \\ \label{kappaH==}
\delta_\kappa {\cal H} &=&  \frac 1{ 2\sqrt{{\cal M}}}{\rm tr} \left(\kappa^q{\mathbf{\Psi}}_q\left( [{\bb X}^i, {\bb P}^i] -4i\{{\mathbf{\Psi}}_p, {\mathbf{\Psi}}_p\}\right) \right) - \frac 1 {\mu^6} \,  \frac {{\cal M}^\prime}{{\cal M}} \,\delta_\kappa {\cal H}\, {\frak H}~, \qquad
\eea
 where  $\frak{H}$ is given in Eq.  \eqref{frakH=} \footnote{To obtain \eqref{kappaH==} one has to use the identity
$\gamma^i_{s(r}\gamma^i_{pq)}\equiv \delta_{s(r}\delta_{pq)}$ and also notice that
 $ {\rm tr} ( {\mathbf{\Psi}}_r\{ {\mathbf{\Psi}}_p,  {\mathbf{\Psi}}_q\})= {\rm tr} ( {\mathbf{\Psi}}_{(r}\{ {\mathbf{\Psi}}_p,  {\mathbf{\Psi}}_{q)}\})$ is completely symmetric with respect to $(rpq)$ indices while   ${\rm tr} ( [(\gamma^i {\mathbf{\Psi}})_q, {\bb X}^i]\,[(\gamma^j {\mathbf{\Psi}})_q, {\bb X}^j])=0$ vanishes.
}. These equations  are solved by \eqref{kappaK=} and \eqref{kappaH=}.

Thus, worldline supersymmetry transformations of the matrix matter fields are given by \eqref{susy=X}-\eqref{susy=Psi} with
\eqref{kappaK=} and \eqref{kappaH=}.

\subsection{Worldline supersymmetry transformations of the worldvolume gauge field}
Taking into account the above results for supersymmetry transformations of the matrix matter fields, we find that the remaining variation of the Lagrangian form \eqref{susy=cLmD0} can be written as
\begin{equation}
\begin{array}{c}
\begin{split}
\mu^6\delta_\kappa {\cal L}_{\text{mD}0} &= {\rm tr} \left(\delta_\kappa {\bb A}\left([{\bb X}^i,{\bb P}^i]- 4i \{ \mathbf{\Psi}_q,\mathbf{\Psi}_q \}\right) \right) + \frac 2 {{\cal M}} \left(1- \frac {{\cal H}} {\mu^6} \, \frac {{\cal M}^\prime}{{\cal M}}\right) E^0\delta_\kappa {\cal H}~+\\
%&~\\
&+ \frac 1 {\sqrt{2{\cal M}}} (E^{1q}-E_q^2)\, \left(  i\delta_\kappa\nu_q -\frac { 4i} {\sqrt{{\cal M}}}\kappa^{q}\, {\cal H}    + \frac 1 {\mu^6} \, \frac 1 {2} \frac {{\cal M}^\prime}{{\cal M}} \, i \nu_q \delta_\kappa {\cal H} \right)~.
\end{split}
\end{array}
\label{susy=cLmD0-1}
\end{equation}
To proceed further, we calculate $ i\delta_\kappa\nu_q $ which reads
\bea
 i\delta_\kappa\nu_q &=& -  \frac {1} {\sqrt{{\cal M}}}\, (\kappa\gamma^i)_q \, {\rm tr} \left({\bb X}^i \, ([{\bb X}^j,{\bb P}^j]- 4i \{{\mathbf{\Psi}}_r,{\mathbf{\Psi}}_r\} ) \right)+ \frac { 4i} {\sqrt{{\cal M}}}\kappa^{q}\, {\cal H} +  \qquad \nonumber \\
% ~\nonumber\\
  &+& \frac 1 {\mu^6} \, \frac {{\cal M}^\prime}{{\cal M}} \,   {\rm tr}\left(4i (\gamma^i {\mathbf{\Psi}})_q  {\bb P}^i +
(\gamma^{ij} {\Psi})_q  [{\bb X}^i, {\bb X}^j]  \right)\, \delta_\kappa {\cal H} - \qquad  \\
%~\nonumber\\
&-& \frac 1 {\mu^6} \, \frac {{\cal M}^\prime}{{\cal M}} \, \, \Delta_\kappa {\cal K}\, {\rm tr} \left({\mathbf{\Psi}}_q ([{\bb X}^i,{\bb P}^i]- 4i \{{\mathbf{\Psi}}_q,{\mathbf{\Psi}}_q \})\right)
 \; \qquad \nonumber
\eea
and substitute it to \eqref{susy=cLmD0-1} arriving at
\bea\label{susy=cLmD0-2}
\mu^6\delta_\kappa {\cal L}_{\text{mD}0}=  {\rm tr} \left(([{\bb X}^i,{\bb P}^i]- 4i \{{\mathbf{\Psi}}_r,{\mathbf{\Psi}}_r\} ) \, \left[\delta_\kappa {\bb A}   + \frac 2 {{\cal M}\sqrt{{\cal M}}}\, E^0\,  (\kappa^q{\mathbf{\Psi}}_q) \frac {\left(1- \frac 1 {\mu^6} \, \frac {{\cal M}^\prime}{{\cal M}}{\cal H}\right)}{\left(1+ \frac 1 {\mu^6} \, \frac {{\cal M}^\prime}{{\cal M}}\, {\frak H}\right)}  \right. \right. \qquad \nonumber \\
%~\nonumber\\
-\frac 1 {\sqrt{2}{\cal M}} \, (E^{1q}-E_q^2)(\gamma^i\kappa)_q \,  {\bb X}^i~+ \nonumber \\
%~\nonumber\\
 \left.\left.+  (E^{1q}-E_q^2)\,  \frac 1 {\mu^6} \, \frac {{\cal M}^\prime}{2{\cal M}\sqrt{2{\cal M}}} \, \left(  -2 \Delta_\kappa {\cal K}\, {\mathbf{\Psi}}_q  + \frac  {\kappa^p{\mathbf{\Psi}}_p} {\sqrt{{\cal M}}}\; \frac {  {\rm tr}\left(4i (\gamma^i {\mathbf{\Psi}})_q  {\bb P}^i +\frac 5 2
(\gamma^{ij} \mathbf{\Psi})_q  [{\bb X}^i, {\bb X}^j]  \right)}{\left(1+ \frac 1 {\mu^6} \, \frac {{\cal M}^\prime}{{\cal M}}\, {\frak H}\right)}\, \right)\right]
 \right)
 . \nonumber \\ {}
\eea
The above expression vanishes if the SU$(N)$ gauge field transforms under worldline supersymmetry as
\bea
\delta_\kappa {\bb A} &=&  - \frac 2 {{\cal M}\sqrt{{\cal M}}}\, E^0\,  (\kappa^q{\mathbf{\Psi}}_q) \frac {\left(1- \frac 1 {\mu^6} \, \frac {{\cal M}^\prime}{{\cal M}}{\cal H}\right)}{\left(1+ \frac 1 {\mu^6} \, \frac {{\cal M}^\prime}{{\cal M}}\, {\frak H}\right)}  +\frac 1 {\sqrt{2}{\cal M}} \, (E^{1q}-E_q^2)(\gamma^i\kappa)_q \,{\bb X}^i~+ \qquad  \nonumber \\
%~\\
& +&  (E^{1q}-E_q^2)\,  \frac 1 {\mu^6} \, \frac {{\cal M}^\prime}{2{\cal M}\sqrt{2{\cal M}}} \, \left(  2 \Delta_\kappa {\cal K}\, {\mathbf{\Psi}}_q  - \frac  {\kappa^p{\mathbf{\Psi}}_p} {\sqrt{{\cal M}}}\; \frac {{\rm tr}\left(4i (\gamma^i \mathbf{\Psi})_q  {\bb P}^i +\frac 5 2
(\gamma^{ij} \mathbf{\Psi})_q  [{\bb X}^i, {\bb X}^j]  \right) }{\left(1+ \frac 1 {\mu^6} \, \frac {{\cal M}^\prime}{{\cal M}}\, {\frak H}\right)} \right)~.  \nonumber
\eea
\end{widetext}
Substituting (\ref{kappaK=}) in it, we arrive after some algebra at Eq. \eqref{susy=A}.


\begin{thebibliography}{99}

%\cite{Dai:1989ua}
\bibitem{Dai:1989ua}
J.~Dai, R.~G.~Leigh and J.~Polchinski,
``New Connections Between String Theories,''
Mod. Phys. Lett. A \textbf{4} (1989), 2073-2083
doi:10.1142/S0217732389002331
%909 citations counted in INSPIRE as of 28 Feb 2022

%\cite{Horava:1989ga}
\bibitem{Horava:1989ga}
P.~Horava,
``Background Duality of Open String Models,''
Phys. Lett. B \textbf{231} (1989), 251-257
doi:10.1016/0370-2693(89)90209-8
%314 citations counted in INSPIRE as of 28 Feb 2022

%\cite{Green:1987sp}
\bibitem{Green:1987sp}
M.~B.~Green, J.~H.~Schwarz and E.~Witten,
``Superstring Theory'' V1, CUP, 1987
%299 citations counted in INSPIRE as of 28 Feb 2022



%\cite{Polchinski:1995mt}
\bibitem{Polchinski:1995mt}
J.~Polchinski,
``Dirichlet Branes and Ramond-Ramond charges,''
Phys. Rev. Lett. \textbf{75} (1995), 4724-4727
doi:10.1103/PhysRevLett.75.4724
[arXiv:hep-th/9510017 [hep-th]].
%2813 citations counted in INSPIRE as of 28 Feb 2022

%\cite{Johnson:2003glb}
\bibitem{Johnson:2003glb}
C.V. Johnson,
``D-branes,''
Cambridge Monographs on Mathematical Physics, CUP, 2003, doi:10.1017/CBO9780511606540.
%52 citations counted in INSPIRE as of 28 Feb 2022


%\cite{Howe:1996mx}
\bibitem{Howe:1996mx}
P.~S.~Howe and E.~Sezgin,
``Superbranes,''
Phys. Lett. B \textbf{390} (1997), 133-142
doi:10.1016/S0370-2693(96)01416-5
[arXiv:hep-th/9607227 [hep-th]].
%149 citations counted in INSPIRE as of 30 Oct 2021



%\cite{Cederwall:1996pv}
\bibitem{Cederwall:1996pv}
M.~Cederwall, A.~von Gussich, B.~E.~W.~Nilsson and A.~Westerberg,
``The Dirichlet super three-brane in ten-dimensional type IIB supergravity,''
Nucl. Phys. B \textbf{490} (1997), 163-178
doi:10.1016/S0550-3213(97)00071-0
[arXiv:hep-th/9610148 [hep-th]].
%293 citations counted in INSPIRE as of 30 Oct 2021

%\cite{Aganagic:1996pe}
\bibitem{Aganagic:1996pe}
M.~Aganagic, C.~Popescu and J.~H.~Schwarz,
``D-brane actions with local kappa symmetry,''
Phys. Lett. B \textbf{393} (1997), 311-315
doi:10.1016/S0370-2693(96)01643-7
[arXiv:hep-th/9610249 [hep-th]].
%315 citations counted in INSPIRE as of 30 Oct 2021

%\cite{Cederwall:1996ri}
\bibitem{Cederwall:1996ri}
M.~Cederwall, A.~von Gussich, B.~E.~W.~Nilsson, P.~Sundell and A.~Westerberg,
``The Dirichlet super p-branes in ten-dimensional type IIA and IIB supergravity,''
Nucl. Phys. B \textbf{490} (1997), 179-201
doi:10.1016/S0550-3213(97)00075-8
[arXiv:hep-th/9611159 [hep-th]].
%418 citations counted in INSPIRE as of 30 Oct 2021

%\cite{Aganagic:1996nn}
\bibitem{Aganagic:1996nn}
M.~Aganagic, C.~Popescu and J.~H.~Schwarz,
``Gauge invariant and gauge fixed D-brane actions,''
Nucl. Phys. B \textbf{495} (1997), 99-126
doi:10.1016/S0550-3213(97)00180-6
[arXiv:hep-th/9612080 [hep-th]].
%353 citations counted in INSPIRE as of 30 Oct 2021

%\cite{Bergshoeff:1996tu}
\bibitem{Bergshoeff:1996tu}
E.~Bergshoeff and P.~K.~Townsend,
``Super D-branes,''
Nucl. Phys. B \textbf{490} (1997), 145-162
doi:10.1016/S0550-3213(97)00072-2
[arXiv:hep-th/9611173 [hep-th]].
%518 citations counted in INSPIRE as of 30 Oct 2021

%\cite{Bandos:1997rq}
\bibitem{Bandos:1997rq}
I.~A.~Bandos, D.~P.~Sorokin and M.~Tonin,
``Generalized action principle and superfield equations of motion for D = 10 D p-branes,''
Nucl. Phys. B \textbf{497} (1997), 275-296
doi:10.1016/S0550-3213(97)00258-7
[arXiv:hep-th/9701127 [hep-th]].
%62 citations counted in INSPIRE as of 30 Oct 2021
%\cite{Witten:1995im}
\bibitem{Witten:1995im}
E.~Witten,
``Bound states of strings and p-branes,''
Nucl. Phys. B \textbf{460} (1996), 335-350
doi:10.1016/0550-3213(95)00610-9
[arXiv:hep-th/9510135 [hep-th]].
%1578 citations counted in INSPIRE as of 30 Oct 2021

%\cite{Tseytlin:1997csa}
\bibitem{Tseytlin:1997csa}
A.~A.~Tseytlin,
``On nonAbelian generalization of Born-Infeld action in string theory,''
Nucl. Phys. B \textbf{501} (1997), 41-52
doi:10.1016/S0550-3213(97)00354-4
[arXiv:hep-th/9701125 [hep-th]].
%594 citations counted in INSPIRE as of 30 Oct 2021

%\cite{Myers:1999ps}
\bibitem{Myers:1999ps}
R.~C.~Myers,
``Dielectric branes,''
JHEP \textbf{12} (1999), 022
doi:10.1088/1126-6708/1999/12/022
[arXiv:hep-th/9910053 [hep-th]].
%1364 citations counted in INSPIRE as of 30 Oct 2021

%\cite{Emparan:1997rt}
\bibitem{Emparan:1997rt}
R.~Emparan,
``Born-Infeld strings tunneling to D-branes,''
Phys. Lett. B \textbf{423} (1998), 71-78
doi:10.1016/S0370-2693(98)00107-5
[arXiv:hep-th/9711106 [hep-th]].
%110 citations counted in INSPIRE as of 31 Oct 2021


%\cite{Sorokin:2001av}
\bibitem{Sorokin:2001av}
D.~P.~Sorokin,
``Coincident (super)Dp-branes of codimension one,''
JHEP \textbf{08} (2001), 022
doi:10.1088/1126-6708/2001/08/022
[arXiv:hep-th/0106212 [hep-th]].
%37 citations counted in INSPIRE as of 31 Oct 2021

%\cite{Drummond:2002kg}
\bibitem{Drummond:2002kg}
J.~M.~Drummond, P.~S.~Howe and U.~Lindstrom,
``Kappa symmetric nonAbelian Born-Infeld actions in three-dimensions,''
Class. Quant. Grav. \textbf{19} (2002), 6477-6488
doi:10.1088/0264-9381/19/24/314
[arXiv:hep-th/0206148 [hep-th]].
%29 citations counted in INSPIRE as of 31 Oct 2021

%\cite{Janssen:2002vb}
\bibitem{Janssen:2002vb}
B.~Janssen and Y.~Lozano,
``On the dielectric effect for gravitational waves,''
Nucl. Phys. B \textbf{643} (2002), 399-430
doi:10.1016/S0550-3213(02)00751-4
[arXiv:hep-th/0205254 [hep-th]].
%41 citations counted in INSPIRE as of 31 Oct 2021


%\cite{Panda:2003dj}
\bibitem{Panda:2003dj}
S.~Panda and D.~Sorokin,
``Supersymmetric and kappa invariant coincident D0-branes,''
JHEP \textbf{02} (2003), 055
doi:10.1088/1126-6708/2003/02/055
[arXiv:hep-th/0301065 [hep-th]].
%22 citations counted in INSPIRE as of 31 Oct 2021




%\cite{Janssen:2002cf}
\bibitem{Janssen:2002cf}
B.~Janssen and Y.~Lozano,
``A Microscopical description of giant gravitons,''
Nucl. Phys. B \textbf{658} (2003), 281-299
doi:10.1016/S0550-3213(03)00185-8
[arXiv:hep-th/0207199 [hep-th]].
%50 citations counted in INSPIRE as of 31 Oct 2021

%\cite{Janssen:2003ri}
\bibitem{Janssen:2003ri}
B.~Janssen, Y.~Lozano and D.~Rodriguez-Gomez,
``A Microscopical description of giant gravitons. 2. The AdS(5) x S**5 background,''
Nucl. Phys. B \textbf{669} (2003), 363-378
doi:10.1016/S0550-3213(03)00532-7
[arXiv:hep-th/0303183 [hep-th]].
%32 citations counted in INSPIRE as of 31 Oct 2021

%\cite{Lozano:2005kf}
\bibitem{Lozano:2005kf}
Y.~Lozano and D.~Rodriguez-Gomez,
``Fuzzy 5-spheres and pp-wave matrix actions,''
JHEP \textbf{08} (2005), 044
doi:10.1088/1126-6708/2005/08/044
[arXiv:hep-th/0505073 [hep-th]].
%23 citations counted in INSPIRE as of 31 Oct 2021



%\cite{Howe:2005jz}
\bibitem{Howe:2005jz}
P.~S.~Howe, U.~Lindstrom and L.~Wulff,
``Superstrings with boundary fermions,''
JHEP \textbf{08} (2005), 041
doi:10.1088/1126-6708/2005/08/041
[arXiv:hep-th/0505067 [hep-th]].
%30 citations counted in INSPIRE as of 31 Oct 2021

%\cite{Howe:2006rv}
\bibitem{Howe:2006rv}
P.~S.~Howe, U.~Lindstrom and L.~Wulff,
``On the covariance of the Dirac-Born-Infeld-Myers action,''
JHEP \textbf{02} (2007), 070
doi:10.1088/1126-6708/2007/02/070
[arXiv:hep-th/0607156 [hep-th]].
%46 citations counted in INSPIRE as of 31 Oct 2021

%\cite{Howe:2007eb}
\bibitem{Howe:2007eb}
P.~S.~Howe, U.~Lindstrom and L.~Wulff,
``Kappa-symmetry for coincident D-branes,''
JHEP \textbf{09} (2007), 010
doi:10.1088/1126-6708/2007/09/010
[arXiv:0706.2494 [hep-th]].
%15 citations counted in INSPIRE as of 31 Oct 2021


%\cite{Bandos:2018ntt}
\bibitem{Bandos:2018ntt}
I.~Bandos,
``Supersymmetric action for multiple D0-brane system,''
JHEP \textbf{11} (2018), 189
doi:10.1007/JHEP11(2018)189
[arXiv:1810.01401 [hep-th]].
%0 citations counted in INSPIRE as of 31 Oct 2021

%\cite{Bandos:2021vrq}
\bibitem{Bandos:2021vrq}
I.~Bandos and U.~D.~M.~Sarraga,
``3D supersymmetric nonlinear multiple D$0$-brane action and 4D counterpart of multiple M-wave system,''
[arXiv:2112.14610 [hep-th]].
%0 citations counted in INSPIRE as of 28 Feb 2022


%\cite{deAzcarraga:1982dhu}
\bibitem{deAzcarraga:1982dhu}
J.~A.~de Azcarraga and J.~Lukierski,
``Supersymmetric Particles with Internal Symmetries and Central Charges,''
Phys. Lett. B \textbf{113} (1982), 170-174
doi:10.1016/0370-2693(82)90417-8
%178 citations counted in INSPIRE as of 01 Nov 2021

%\cite{deAzcarraga:1982njd}
\bibitem{deAzcarraga:1982njd}
J.~A.~de Azcarraga and J.~Lukierski,
``Supersymmetric Particles in $N=2$ Superspace: Phase Space Variables and Hamiltonian Dynamics,''
Phys. Rev. D \textbf{28} (1983), 1337
doi:10.1103/PhysRevD.28.1337
%85 citations counted in INSPIRE as of 01 Nov 2021

%\cite{Siegel:1983hh}
\bibitem{Siegel:1983hh}
W.~Siegel,
``Hidden Local Supersymmetry in the Supersymmetric Particle Action,''
Phys. Lett. B \textbf{128} (1983), 397-399
doi:10.1016/0370-2693(83)90924-3
%439 citations counted in INSPIRE as of 01 Nov 2021

%\cite{Sorokin:1988jor}
\bibitem{Sorokin:1988jor}
D.~P.~Sorokin, V.~I.~Tkach and D.~V.~Volkov,
``Superparticles, Twistors and Siegel Symmetry,''
Mod. Phys. Lett. A \textbf{4} (1989), 901-908
doi:10.1142/S0217732389001064
%232 citations counted in INSPIRE as of 05 Sep 2021

%\cite{Sorokin:1989jj}
\bibitem{Sorokin:1989jj}
D.~P.~Sorokin,
``Double supersymemtric particle theories,''
Fortsch. Phys. \textbf{38} (1990), 923-943
ITF-89-3E.
%41 citations counted in INSPIRE as of 26 Feb 2022



%\cite{Bandos:2000tg}
\bibitem{Bandos:2000tg}
I.~A.~Bandos,
``Super D0-branes at the endpoints of fundamental superstring: An Example of interacting brane system,'' In: ''Proceedings, International Workshop on Supersymmetries and Quantum Symmetries (SQS'99): Moscow, Russia, July 27-31, 1999''. JINR, Dubna, 2000  [arXiv:hep-th/0001150 [hep-th]].
%4 citations counted in INSPIRE as of 11 Nov 2021



%\cite{Bandos:2012jz}
\bibitem{Bandos:2012jz}
I.~A.~Bandos,
``Action for the eleven dimensional multiple M-wave system,''
JHEP \textbf{01} (2013), 074
doi:10.1007/JHEP01(2013)074
[arXiv:1207.0728 [hep-th]].
%5 citations counted in INSPIRE as of 31 Oct 2021

%\cite{Bandos:2013uoa}
\bibitem{Bandos:2013uoa}
I.~A.~Bandos and C.~Meliveo,
``Covariant action and equations of motion for the eleven dimensional multiple M0-brane system,''
Phys. Rev. D \textbf{87} (2013) no.12, 126011
doi:10.1103/PhysRevD.87.126011
[arXiv:1304.0382 [hep-th]].
%4 citations counted in INSPIRE as of 31 Oct 2021

%\cite{Bandos:1990ji}
\bibitem{Bandos:1990ji}
I.~A.~Bandos,
``Superparticle in Lorentz harmonic superspace,''
Sov. J. Nucl. Phys. \textbf{51} (1990), 906-914
[{\em Yad. Fiz.} {\bf 50} (1989)
%N9. P.
893-899 in Russian]
%99 citations counted in INSPIRE as of 01 Mar 2022

\bibitem{Igor+Unai=in-prep}
Igor Bandos and Unai D.M. Sarraga, in preparation.

%\cite{Delduc:1991ir}
\bibitem{Delduc:1991ir}
F.~Delduc, A.~Galperin and E.~Sokatchev,
``Lorentz harmonic (super)fields and (super)particles,''
Nucl. Phys. B \textbf{368} (1992), 143-171
doi:10.1016/0550-3213(92)90201-L
%81 citations counted in INSPIRE as of 04 Nov 2021


%\cite{Galperin:1991gk}
\bibitem{Galperin:1991gk}
A.~S.~Galperin, P.~S.~Howe and K.~S.~Stelle,
``The Superparticle and the Lorentz group,''
Nucl. Phys. B \textbf{368} (1992), 248-280
doi:10.1016/0550-3213(92)90527-I
[arXiv:hep-th/9201020 [hep-th]].
%85 citations counted in INSPIRE as of 04 Nov 2021

%\cite{Sokatchev:1985tc}
\bibitem{Sokatchev:1985tc}
E.~Sokatchev,
``Light Cone Harmonic Superspace and Its Applications,''
Phys. Lett. B \textbf{169} (1986), 209-214
doi:10.1016/0370-2693(86)90652-0
%136 citations counted in INSPIRE as of 12 May 2022



\end{thebibliography}
\end{document}